\documentclass[twocolumn,showpacs,preprintnumbers,amsmath,amssymb]{revtex4}
\usepackage{graphicx}% Include figure files
\usepackage{dcolumn}% Align table columns on decimal point
\usepackage{bm}% bold math
\DeclareMathAlphabet{\mathpzc}{OT1}{pzc}{m}{it}

\begin{document}

\title{Inflationary phase and role of dark energy: Revisited}

\author{Indranath Bhattacharyya}
\email{ i\_bhattacharyya@hotmail.com}
\affiliation{Department of Mathematics, Barasat Government College, Barasat, Kolkata 700124, West Bengal, India}

\author{Saibal Ray}
\email{saibal@associates.iucaa.in}
\affiliation{Department of Physics, Government College of Engineering and Ceramic Technology, Kolkata 700010, West Bengal, India}

\author{Prasenjit Paul}
\email{prasenjit071083@gmail.com}
\affiliation{Department of Physics, Government College of Engineering and Ceramic Technology, Kolkata 700010, West Bengal, India}

\date{\today}

\begin{abstract}
The inflationary phase of the Universe is explored by proposing a toy model related to the scalar field, termed as {\it inflaton}.
The potential part of the energy density in the said era is assumed to have a constant vacuum energy density part and a
variable part containing the inflaton. The prime idea of the proposed model constructed in the framework of the closed Universe
is based on a fact that the inflaton is the root cause of the orientation of the space. According to this model the expansion of
the Universe in the inflationary epoch is not approximately rather exactly exponential in nature and thus it can solve some of the
fundamental puzzles, viz. flatness as well as horizon problems. It is also predicted that the constant energy density part in the potential may be associated to the dark energy, which is eventually different from the vacuum energy, at least in the inflationary phase of the Universe. However, the model keeps room for the end of inflationary era. 
\end{abstract}


\maketitle

\section{Introduction}
The cosmology based on an elegant idea that our Universe was
started to be expanded from a spacetime singularity, known as Big
Bang cosmology. The discovery of the Cosmic Microwave Background
~\cite{Gammow} abandoned the idea of the Steady State theory of the
Universe~\cite{Bondi,Hoyle1,Hoyle2}. Moreover, the data observed
in the `Supernova Cosmology Project (SCP)'
~\cite{Perlmutter1,Perlmutter2} and the `High-z Supernova Search
Team (HST)'~\cite{Riess,Schmidt} show that presently the Universe
is not only expanding but also accelerating. Both the groups
concluded that their results are sensitive to a linear combination
of $0.8 \Omega_{M} - 0.6 \Omega_{\Lambda}$ (SCP) and $1.4
\Omega_{M} - \Omega_{\Lambda}$ (HST), where $\Omega_{M}$ and
$\Omega_{\Lambda}$ being respectively the cosmic matter and vacuum
energy densities. The negative sign in the linear combination
signifies that the matter and vacuum energy have opposite effects
on the cosmological acceleration. A positive vacuum energy causes
to accelerate the expansion, while the matter tends to slow it
down. The high density vacuum energy eventually wins over the
matter and makes the Universe to be expanded with an acceleration.
Such vacuum energy, characterized by the negative pressure $p
\simeq -\rho$, is interpreted as dark energy, constituting $73$
percent of the content of the present Universe~\cite{Sahni}. 
The energy associated with primordial inflation is around a few
TeV having extremely long duration and hence dark energy be 
a natural consequence of inflationary paradigm at the electroweak energy scale~\cite{Ringeval}.

During early age the Universe was solely filled up with this
vacuum energy. The scale factor $a(t)$ grew exponentially so that
the ratio of the vacuum energy density would rapidly be
approaching towards the critical density at this era. Such
phenomenon of exponential expansion is called {\it inflation}. It
was Guth~\cite{Guth1,Guth2} who first proposed such inflationary model 
to fix up `monopole problem'. It was soon
observed that the inflation could solve other long-standing
problems, such as `flatness problem' and `horizon problem' also.
Guth proposed that the scalar fields might get caught in the local
minimum of some potential, and then rolled towards a true minimum
of the potential. Kazanas~\cite{Kazanas} suggested that an
exponential expansion could eliminate the particle horizon.
According to Sato~\cite{Sato} exponential expansion would be able
to eliminate domain walls, another kind of exotic relic. Einhorn
and Sato~\cite{Einhorn} argued that their model could solve
`magnetic monopole puzzle'. 

However, it was then realized that Guth's model
of inflation had fatal problem as the transition from super cooled
initial `false vacuum'~\cite{Blome,Davies,Hogan,Kaiser} to the
lower energy `true vacuum' cannot occur everywhere simultaneously
~\cite{Hawking,Weinberg}. The Guth version of inflation theory was
then necessarily replaced by a new type `slow roll inflation'
theory~\cite{Linde,Albrecht}. A scalar field, called
`inflaton' was introduced to explain such phenomenon. Coleman and
Weinberg~\cite{Coleman} introduced the symmetry breaking mechanism
to the new inflation model. In the inflation theory Linde
~\cite{Linde3} added a new dimension by introducing the `chaotic
inflation' in which initially one or more inflaton fields vary in
a random manner with positions. Linde~\cite{Linde4,Linde5} also
proposed the theory of `hybrid inflation' theories by introducing
more than one scalar fields.

In cosmology essentially inflation is nothing but a rapid
exponential expansion of the early Universe by a factor of at
least $10^{78}$ in volume, driven by negative pressure. The
detailed particle physics mechanism behind such phenomenon is
still unknown to the researchers. It is also a super cooled
expansion, when the temperature drops down from $10^{27}$ K to
$10^{22}$ K, although exact drop of temperature is model
dependent. Such relatively low temperature is maintained
throughout the entire inflationary phase and at the end of
inflation the reheating occurs to attain the pre-inflationary
temperature.

In the present article our motivation is to propose a new kind of
inflation model based on the scalar potential with decreasing
kinetic energy. The energy density is assumed to consist of a
large constant part along with a small slowly varying part, termed
as vacuum energy density. The constant part would generate an
exponential expansion, whereas the variable part of the energy
density participates in the dynamics of inflation. In a nutshell,
therefore, we are strongly motivated by the idea that during
inflationary age the expansion is approximately exponential in
nature. In the ongoing work, basically we investigate role of 
dark energy in the inflationary model. 

The study is organized as follows: 
We provide the basic mathematical formulation including 
the Einstein field equations in Sec.~2. In Sec.~3 we have 
presented physical explanation of the proposed model, 
whereas some case studies have been featured in Sec.~4. 
The last Sec.~5 is devoted to provide some concluding remarks 
along with salient features.

\section{Mathematical formulation}
We consider a scalar field $\phi$, known as the inflaton, along
with a potential $V(\phi)$ to take part to vary the vacuum energy.
At this stage of expansion the Universe is assumed to be filled up
with such vacuum energy. The expression of the vacuum energy
density as well as vacuum pressure in presence of a spatially
homogeneous inflaton $\phi$ in the Robertson-Walker spacetime
becomes~\cite{Weinberg2}
\begin{equation}
  \rho=\frac{1}{2}\dot{\phi}^{2}+V(\phi),~~~
  p=\frac{1}{2}\dot{\phi}^{2}-V(\phi),
\end{equation}
which must be derived from the expression of the
generalized energy-momentum tensor
\begin{equation}
T^{\mu\nu}_{\phi}=-g^{\mu\nu}\left[\frac{1}{2}g^{\rho\sigma}\frac{\partial
\phi} {\partial x^{\rho}}\frac{\partial \phi}{\partial
x^{\sigma}}+V(\phi)\right] +g^{\mu\rho}g^{\nu\sigma}\frac{\partial
\phi}{\partial x^{\rho}}\frac{\partial \phi}{\partial x^{\sigma}}.
\end{equation}

Such energy-momentum tensor is slightly different from that of the
constant vacuum energy $T^{\mu\nu}$, which is proportional to 
$g^{\mu\nu}$ in the general coordinate system according to the
Lorentz invariance condition. The proportionality of $T^{\mu\nu}$
to $g^{\mu\nu}$ corresponds to the condition of negative pressure
as $p=-\rho$, but by the introduction of inflaton such condition
is deviated as neither pressure nor density remains constant.

Now the important issue is what should be the expression of
$V(\phi)$. Let us, as a check, consider the expression of
potential in the following form:
\begin{equation}
V(\phi)=\frac{1}{2}m^{2}\phi^{2}+\rho_{\Lambda}, \label{potential}
\end{equation}
where $\rho_{\Lambda}$ is the constant vacuum density.

It looks like the potential of the massive scalar field along with
this constant term $\rho_{\Lambda}$, which should be much larger than
the term $\frac{1}{2}m^{2}\phi^{2}$. The idea behind to introduce
such potential is that it should be analogous to the vacuum energy
density if there is no inflaton field at all.

Now, let us recall the Friedmann equation in presence of inflaton,
which yields following two equations:
\begin{equation}
\ddot{\phi}+3H\dot{\phi}+m^{2}\phi=0,
\end{equation}

\begin{equation}
\frac{8\pi
G}{3H^{2}}\left[\frac{\dot{\phi}^{2}}{2}+\frac{m^{2}\phi^{2}}{2}+
\rho_{\Lambda}\right]-1=\frac{k}{a^{2}H^{2}},
\end{equation}
where $H$ represents the Hubble parameter and $m$ is the mass of
inflaton field.

It is already mentioned that we are strongly motivated by the idea
that during inflationary age the expansion is approximately
exponential in nature. The presence of inflaton makes the Universe
to expand very fast, nearly exponentially. Therefore, one may
assume that the $\phi$ field is inversely proportional to the
scale factor $a$. Such proportionality relation may be expressed
as
\begin{equation}
a\phi=c_{k}=\sqrt{\frac{3k}{4\pi c_{p}^{2}}},
\end{equation}
where $c_{p}$ stands for the ratio of the mass of the inflaton to
the Planck mass. The above relation implies that the expansion of
the Universe yields the decay of inflaton field in the same rate,
although it is restricted in the inflationary era. Now, using the
proportionality relation given by the above equation the Hubble
parameter can be calculated from Eq. (5) as
\begin{equation}
H=H_{I}\left[1+\frac{\dot{\phi}^{2}}{2\rho_{\Lambda}}\right]^{\frac{1}{2}},
\end{equation}
where
\begin{equation}
H_{I}=\sqrt{\frac{8\pi G \rho_{\Lambda}}{3}}.
\end{equation}

If $\rho_{\Lambda}\gg \dot{\phi}^{2}$ the Hubble parameter can be approximated as
\begin{equation}
H\approx H_{I}+
\dot{\phi}^{2}\left(\frac{H_{I}}{2\rho_{\Lambda}}\right).
\end{equation}

Let us now try to solve Eq. (4) with the aid of Eq. (9).
Neglecting the second degree term of $\dot{\phi}$ one can solve
the equation to obtain the expression of inflaton as

\begin{equation}
\phi=e^{-\frac{3H_{I}}{2}t}\left[Ae^{-\frac{\sqrt{9H_{I}^{2}-4m^{2}}}{2}t}+Be^{\frac{\sqrt{9H_{I}^{2}-4m^{2}}}{2}t}\right].
\end{equation}

We would like to obtain the analytical expression of the scale
factor to realize the nature of the expansion in the inflationary
age. By means of Eqs. (6), (7) and (8) such expression can be
calculated as
\begin{eqnarray}
\left(a+\sqrt{a^{2}-\frac{k}{m^{2}}}\right)e^{-\frac{\sqrt{a^{2}-\frac{k}{m^{2}}}}{a}} \\ \nonumber =
\left(a_{I}+\sqrt{a_{I}^{2}-\frac{k}{m^{2}}}\right)e^{H_{I}t-\frac{\sqrt{a_{I}^{2}-\frac{k}{m^{2}}}}{a_{I}}},
\end{eqnarray}
where $a_{I}$ is the initial value of $a$, i.e., the value of
scale factor  at the beginning of inflationary era. The value of
$a$, smaller than $a_{I}$, corresponds to the quantum age. If $m
a_{I}\gg 1$ the equation (11) gives
\begin{equation}
a\approx a_{I}e^{H_{I}t},
\end{equation}
showing the approximate nature of exponential expansion.

To have a simplified as well as approximate expression of $\phi$ one can
exploit the proportionality relation given by Eq. (6) as well as
approximate expression of $a$ in Eq. (12) and find that the $\phi$
is approximately proportional to $e^{H_{I}t}$. This is possible if
in Eq. (10) it is assumed $A=0$ and $B=\phi_{I}$ (where
$\phi_{I}$ is the value of $\phi$ at the beginning of inflation phase)
or vise-versa. In addition to that the following relation is to be
satisfied, i.e.
\begin{equation}
H_{I}=\frac{m}{\sqrt 2}.
\end{equation}

This is an important relation showing that $m$ is neither
imaginary nor zero because of the expression of $H_{I}$ being
positive real quantity. Again, according to Eq.~(6) the $k$ is
proportional to $m^{2}$ and therefore the model proposed here is
restricted only for the closed type of Universe. It is also to be
noted that Eq. (6) shows a relation between the spacetime geometry
and inflaton field, which readily solves the flatness problem and
yields an approximate exponential expansion. The flatness
condition $|\frac{V'(\phi)}{V(\phi)}|\ll \sqrt{16\pi G}$
~\cite{Weinberg2} gives an estimate of $\phi$. Utilizing this
flatness condition and also the condition given by Eq. (10)
one can estimate
\begin{equation}
\frac{|\phi_{I}|}{m_{p}}\ll 0.24
\end{equation}
where $m_{p}$ represents the Planck mass.

In this model we must keep provision for the Universe to get out
of the inflationary era and to begin the radiation dominated
regime. In the present model $\phi$
as well as $\dot{\phi}$ diminish exponentially with time and 
hence $\rho_a$. Eventually $\rho_a$ vanishes when $t$ is large enough.
This is quite uncomfortable situation as it implies the
inflation could take pretty long time to be completed.
To get rid of this situation it may be assumed that $\rho_a$ 
need not vanish rather would reach to a minimum value $\rho_{min}$ 
at the end of the inflationary phase. After that $\rho_a$ 
decays into standard model particles including the
electromagnetic radiation, resulting the ‘radiation dominated’
phase. It is to be noted that all the particles at
this stage remain massless. Some of these may get mass
through spontaneous symmetry breaking. The process of
decaying $\rho_a$ into the particles might take place through
parametric resonance \cite{Kofman}, which is a mechanical excitation
and oscillation at certain frequencies. 

The above issue can also be handled 
in a mathematical framework so that the inflation could end up. 
The total density $\rho$ of the Universe may be expressed as
\begin{equation}
\rho= \rho_{a}+\rho_{\Lambda},
\end{equation}
where
\begin{equation}
\rho_{a}=\frac{1}{2}(\dot{\phi}^{2}+m^{2}\phi^{2}).
\end{equation}

From the energy equation of the cosmology, given by
\begin{equation}
\dot{\rho}=-\frac{3\dot{a}}{a}(p+\rho),
\end{equation}
with $$ p=w\rho, $$ it can easily be obtained $w= - \frac{1}{3}$,
if the density varies as $a^{-2}$. If the density would comprise
of only constant vacuum density $\rho_{\Lambda}$, then $w=-1$. In
the Eq. (14) $\rho_{a}$ is the varying density, whereas
$\rho_{\Lambda}$ is the constant vacuum density. Now one can treat
such varying density as a state and the pressure-density relation
may be realized as
\begin{equation}
p_{a}=W\rho_{a}=w\rho_{a}
\end{equation}
where $W$ is an operator which brings different phases of the
universe and $w$ is the corresponding eigen value. In the
inflationary phase of the Universe the eigen value $w= -
\frac{1}{3}$. Now the operator $W$ operating on $\rho_{a}$ brings
a change so that the eigen value $w$ becomes $\frac{1}{3}$,
leading to the end of inflation. It can be calculated that when
$w= \frac{1}{3}$ the density $\rho_{a}$ is proportional to
$a^{-4}$ and thus the radiation age begins. At the radiation phase
the varying density energy becomes
\begin{equation}
\rho_{a}=\rho_{R}\left(\frac{a}{a_{R}}\right)^{-4},
\end{equation}
where $\rho_{R}$ and $a_{R}$ represent the radiation density and
the scale factor, respectively, at the moment when inflation ends
and radiation era begins. Such a radiation density $\rho_{R}$ can be
estimated as
\begin{equation}
\rho_{R}\approx \left(\frac{3k}{8\pi G}\right)\frac{1}{a_{R}^{2}}.
\end{equation}

Now a question may arise what will be the nature of $\phi$ and
$\dot{\phi}$ at the end of inflation. Relative to the time scale
of inflationary age both of $\phi$ and $\dot{\phi}$ vanish
asymptotically and thus the potential $V$ reaches to a true
vacuum. But in the global scale of time period $\phi$ should have
a minimum value beyond which no inflaton field exists and the
inflation era gradually exits to the radiation age. This minimum
$\phi$ is given by
\begin{equation}
\phi_{min}=\frac{c_{k}}{a_{R}}.
\end{equation}

This model does not deal with the dynamics of the Universe in the
radiation age. It only addresses the issue of possible deadlock at
the end of inflation.

The number of e-foldings can be calculated as
\begin{equation}
N=H_{I}\left[T-\left(\frac{H_{I}\phi_{I}^{2}}{4\rho_{\Lambda}}\right)e^{-2H_{I}T}\right],
\end{equation}
where $T$ is the total time period for which the inflation lasts.

Thus it can be predicted for a large period that the number of
e-foldings are directly proportional to the time period through
which the inflation elapses. To have a large number of e-foldings
therefore the time period of the inflationary phase is to be large
enough. The minimum amount of inflation required to solve the
various cosmological problems is about 70 e-foldings, i.e. an
expansion by a factor of $10^{30}$. The total number of e-foldings
inflation described by this model will exceed the number depending
on the period of inflation as well as mass of the inflaton field.

\section{Physical explanation}
To this end, let us critically discuss how the present model is
advantageous compared to the earlier models. It has already been
mentioned that Guth's model of inflation~\cite{Guth1,Guth2}
had problem that the transition from the `false vacuum'
~\cite{Blome,Davies,Hogan,Kaiser} to the `true vacuum' does not
occur everywhere at a time. Linde's version could able to get rid
of such drawback and well explain `flatness' and `horizon'
problem. Likewise Linde's model~\cite{Linde} our model
proposes an approximate nature of such exponential burst.

In the framework of the proposed model an important relation is
deduced in Eq. (13). The left hand side of the said equation
contains the terms representing the dynamics of the expansion,
whereas the right hand side stands for the space-time geometry.
Followed by Eq. (13) it is quite evident that $k$ is not only
non-zero but it is also a positive, interpreting the orientation
of the Universe is closed. In Eq. (5) the term $k$ is
cancelled out, which indicates that the expansion of the Universe
is independent of the orientation of the space. In the closed
Universe the massive inflaton field makes the Universe flat in
that stage. Thus the presence of inflaton field not only fix the
`flatness problem', but it may also be interpreted that the origin
of the inflaton is solely responsible for the orientation of
space.

The density $\rho$ of the Universe at this stage can be divided
into two parts: the variable part $\rho_a$ and the constant
part $\rho_{\Lambda}$. This constant part $\rho_{\Lambda}$ does not take part in
the dynamical change of the Universe. Now one can interpret
such energy $\rho_{\Lambda}$ as the widely discussed ‘dark energy’.
The dark energy identified in this manner is different
from the vacuum energy consisting of both $\rho_a$ and 
$\rho_{\Lambda}$ in the inflation era. With a minimal prediction one must
conclude that the dark energy, arising by the introduction
of cosmological constant, is different from vacuum
energy even beyond the inflationary age, if quintessence
effect~\cite{ratra} has not been taken into account. However, Basilakos, Lima and Sola~\cite{BLS}
have made an attempt at elleviating fundamental cosmic puzzles via a dynamic $\Lambda$ 
which may have definite role for “gracefully” exits from inflation to a radiation phase followed by
dark matter and vacuum regimes, and, finally, evolves to a late-time de Sitter phase.

\section{Comparison with other inflationary models} 
In this section we compare our toy model with two recent inflationary models, 
viz. intermediate inflation and logamediate inflation~\cite{barrow1,barrow2,barrow3,Herrera}, 
in the light of recent obsevational data~\cite{wmap,planck1,planck2}.

 Using the potential given in Eq. (\ref{potential}), we can obtain the slow roll parameter as
 \begin{equation}
 \epsilon(\phi)=\frac{M_p^2}{2}\left[\frac{V'(\phi)}{V(\phi)}\right]^2=\frac{M_p^2}{2}\left[\frac{m^2\phi}{\rho_{\Lambda}+\frac{1}{2}m^2\phi^2}\right]^2,  \label{0-5}
\end{equation}

\begin{equation}
 \eta(\phi)=-M_p^2\left[\frac{V''(\phi)}{V(\phi)}\right]=-\left[\frac{M_p^2m^2}{\rho_{\Lambda}+\frac{1}{2}m^2\phi^2}\right].  \label{0-6}
\end{equation}

Generally, inflationary models are characterized by the scalar spectral index $n_s$ and the
tensor-to-scalar ratio $r$~\cite{starobinsky1,starobinsky2}. The scalar spectral index ($n_s$) and the tensor-to-scalar ratio ($r$)
of primordial spectrum which are related to the slow-roll parameters ($\epsilon, \eta$), given by
\begin{equation}
n_s-1\approx 2\eta-6\epsilon, \label{0-9}
\end{equation}

\begin{equation}
r\approx 16\epsilon. 
\end{equation} 

\begin{figure}
\begin{center}
\includegraphics[width=1.20\linewidth]{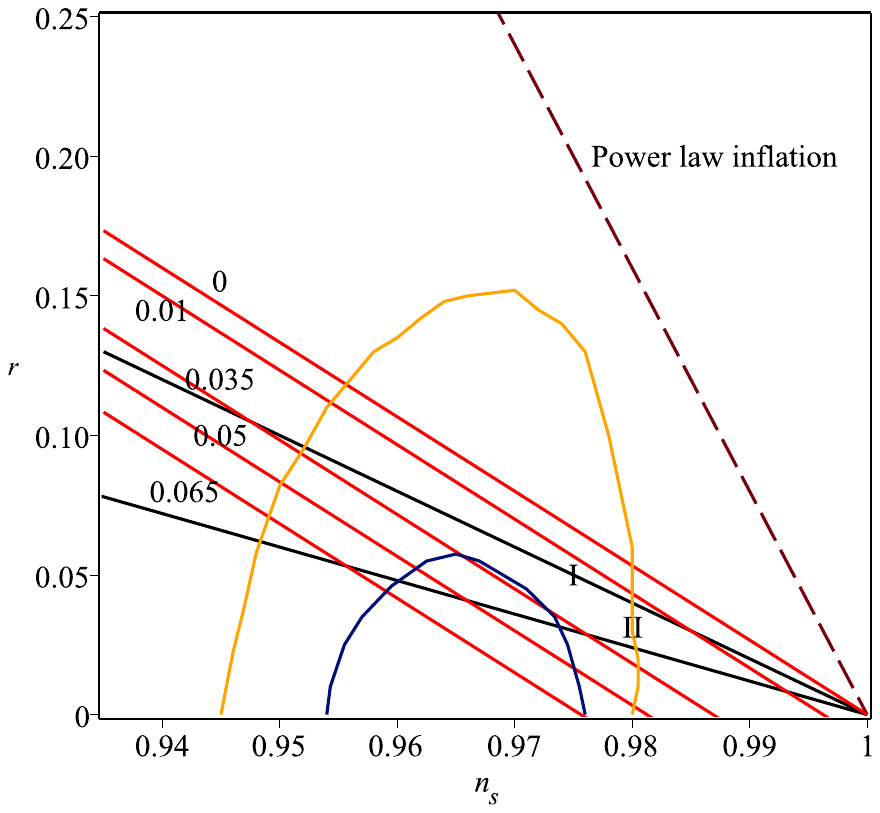}
\vspace{-6cm}
\caption{The plot of $r$ versus $n_s$ for different values of $C$ (red lines). Outer contour correponds to the Planck + WMAP-9 + BAO data~\cite{wmap,planck1} whereas inner contour correponds to the Planck + BICEP2 + Keck Array data~\cite{planck2}. The dashed line corresponds to power-law inflationary models with exponential potentials whereas the black lines I and II correspond to intermediate and logamediate inflation model respectively.}
\end{center}
\end{figure}

Assuming $\rho_{\Lambda}$ to be large and constant as indicated earlier for the present model, we can write
 \begin{equation}
 r \approx \frac{-8}{3}(n_s-1)-C,
 \label{relation2}
 \end{equation}
where $C=\frac{16M_p^2m^2}{3\rho_{\Lambda}}$. Hence it is possible to vary $C$ in the range $C~\in$ [0,0.065] so that our toy model fits well in $\{r, n_s\}$ space with the observational data. 

Now, in the case of intermediate inflation, we have $a = exp(At^f)$, where $A > 0$ and $1 > f > 0$ are constants. In Fig. 1 we take $A=0.3$ and f=0.39~\cite{barrow2,Herrera}. On the other hand, in the case of logamediate inflation we have $a = exp[A(\ln t)^{\lambda}]$, where $A > 0$ and $\lambda > 1$ are constants. In Fig. 1 we take, $A=2.1 \times 10^{-2}$ and $\lambda$=2~\cite{barrow3}.

Fig. 1 shows that our model agrees fairly well with the observational data along with these highly modified inflationary models.

\section{Conclusion}
To summarize, the basic philosophy behind the present paper is to
explain the nature of exponential expansion in general.
However, specifically the investigation provides the following
novel features:

(i) The model shows that in the inflationary era the Hubble constant,
an approximate constant and hence a variable, evolves out of the inflaton mass.

(ii) The model keeps room for the end of inflation to avoid the possible deadlock.

(iii) It can also be predicted that for a large time period the number of e-foldings are
directly proportional to the time period through which the
inflation elapses. In other words, to have a large number of e-folding the time
period of the inflationary phase is to be large enough.

(iv) In the treatment we have divided the density $\rho$ of the Universe into two parts: 
the variable part $\rho_a$ and the constant part $\rho_{\Lambda}$. Though the constant part $\rho_{\Lambda}$ does not take part in the dynamical change of the Universe, however it can be interpreted as the widely discussed ‘dark energy’~\cite{ratra} responsible for inflation 
through it's repulsive nature.

(v) It has been shown that even in such a simple toy model the values 
of the primordial tilt and the tensor-to-scalar ratio are in very good 
agreement with the observational data and comparable with other modified inflationary models.

It has already been shown that the proposed model can explain the
flatness as well as the horizon problems. However, there is a
great issue of monopole. It seems that at this stage monopole
problem cannot be tackled by the model as it is fixed through the
process of reheating after the inflationary phase. This is
therefore beyond purview of the present work.

\section*{Acknowledgement}
SR is thankful to support from Inter-University Centre for
Astronomy and Astrophysics, Pune, India under which a part of this
work was carried out.


\begin{thebibliography}{99}

\bibitem{Gammow} G. Gammow, ``Expanding Universe and the Origin of Elements'', {\it Physical Review}, vol. 70, p. 572, 1946.

\bibitem{Bondi} H. Bondi and T. Gold, ``The Steady-State Theory of the Expanding Universe'', {\it Monthly Notices of the Royal Astronomical Society}, vol. 108, p. 252, 1948.

\bibitem{Hoyle1} F. Hoyle, ``A New Model for the Expanding Universe'', {\it Monthly Notices of the Royal Astronomical Society}, vol. 108, p. 372, 1948.

\bibitem{Hoyle2} F. Hoyle, ``On the Cosmological Problem'', {\it Monthly Notices of the Royal Astronomical Society}, vol. 109, p. 365, 1949.

\bibitem{Perlmutter1} S. Perlmutter et al., ``Measurements of~$\Omega$ and $\Lambda$ from 42 High-Redshift Supernovae'', {\it Astrophysical Journal}, vol. 517, p. 565, 1999.

\bibitem{Perlmutter2} S. Perlmutter et al., ``Discovery of a supernova 	explosion at half the age of the Universe'', {\it Nature}, vol. 392, p. 51, 1998.

\bibitem{Riess} A.G. Riess et al., ``Observational Evidence from Supernovae for an Accelerating Universe and a Cosmological Constant'', {\it Astronomical Journal}, vol. 116, p. 1009, 1998.

\bibitem{Schmidt} B. Schmidt et al., ``The High-Z Supernova Search: Measuring Cosmic Deceleration and Global Curvature of the Universe Using Type Ia Supernovae'', {\it Astrophysical Journal}, vol. 507, p. 46, 1998.

\bibitem{Sahni} V. Sahni, ``The Physics of the Early Universe: Dark Matter and Dark Energy'', {\it Lecture Notes on Physics}, vol. 653, p. 141, 2004.

\bibitem{Ringeval} C. Ringeval, ``Dark energy from inflation'', {\it Journal of Physics: Conference Series}, vol. 485, 012023, 2014. 

\bibitem{Guth1} A. Guth, {\it The Inflationary Universe}, Reading, Massachusetts, 1997.

\bibitem{Guth2} A. Guth, ``Inflationary universe: A possible solution to the horizon and flatness problems'', {\it Physical Review D}, vol. 23, p. 347, 1981.

\bibitem{Kazanas} D. Kazanas, ``Dynamics of the universe and spontaneous symmetry breaking'', {\it Astrophysical Journal}, vol. 241, L59, 1980.

\bibitem{Sato} K. Sato, ``Cosmological baryon-number domain structure and the first order phase transition of a vacuum'', {\it Physics Letters B}, vol. 33, p. 66, 1981.

\bibitem{Einhorn} M.B. Einhorn and K. Sato, ``Monopole production in the very early universe in a first-order phase transition'', {\it Nuclear Physics B}, vol. 180, p. 385, 1981.

\bibitem{Blome} J.J. Blome and W. Priester, ``Vacuum energy in a Friedmann-Lemaître cosmos'', {\it Naturwissenschaften}, vol. 71, p. 528, 1984.

\bibitem{Davies} P.C.W. Davies, ``Vacuum energy in a Friedmann-Lemaître cosmos'', {\it Physical Review D}, vol. 30, p. 737, 1984.

\bibitem{Hogan} C. Hogan, ``Cosmic strings and galaxies'', {\it Nature}, vol. 310, p. 365, 1984.

\bibitem{Kaiser} N. Kaiser and A. Stebbins, ``Microwave anisotropy due to cosmic strings'', {\it Nature}, vol. 310, p. 391, 1984.

\bibitem{Hawking} S.W. Hawking, I.G. Moss and J.M. Stewart, ``Bubble collisions in the very early universe'', {\it Physical Review D}, vol. 26, p. 3681, 1982.

\bibitem{Weinberg} A.H. Guth and E.J. Weinberg, ``Could the universe have recovered from a slow first-order phase transition?'' {\it Nuclear Physics B}, vol. 212, p. 321, 1983.

\bibitem{Linde} A.D. Linde, ``A new inflationary universe scenario: A possible solution of the horizon, flatness, homogeneity, isotropy and primordial monopole problems'', {\it Physics Letters B}, vol. 108, 389 (1982); ``Coleman-Weinberg theory and the new inflationary universe scenario'', vol. 114, p. 431, 1982.

\bibitem{Albrecht} A. Albrecht and P. Steinhardt, ``Cosmology for Grand Unified Theories with Radiatively Induced Symmetry Breaking'', {\it Physical Review Letters}, vol. 48, p. 1220, 1982.

\bibitem{Coleman} S. Coleman and E. Weinberg, ``Radiative Corrections as the Origin of Spontaneous Symmetry Breaking'', {\it Physical Review D}, vol. 7, p. 1888, 1973.

\bibitem{Linde3} A.D. Linde, ``Chaotic inflation'', {\it Physics Letters B}, vol. 129, p. 177, 1983.

\bibitem{Linde4} A.D. Linde, ``Axions in inflationary cosmology'', {\it Physics Letters B}, vol. 259, p. 38, 1991.

\bibitem{Linde5} A.D. Linde, ``Hybrid inflation'', {\it Physical Review D}, vol. 49, p. 748, 1994.

\bibitem{Weinberg2} S. Weinberg, {\it Cosmology}, Chapter 4, Oxford Univ., New
York, 2008.

\bibitem{Kofman} L. Kofman, A.D. Linde and A. Starobinsky, ``Reheating after Inflation'', {\it Physical Review Letters}, vol. 73, p. 3195, 1994.

\bibitem{ratra} B. Ratra and P.J.E. Peebles, ``The Cosmological Constant and Dark Energy'', {\it Reviews of Modern Physics}, vol. 75, p. 559, 2003.

\bibitem{BLS} S. Basilakos, J.A.S. Lima and J. Sola, ``From inflation to dark energy through a dynamical Lambda: an attempt at alleviating fundamental cosmic puzzles'', {\it International Journal of Modern Physics D}, vol. 22, 1342008, 2013.

\bibitem{barrow1} J.D. Barrow and P. Saich, ``The behaviour of intermediate inflationary universes'', {\it Physics Letters B}, vol. 249, p. 406, 1990.

\bibitem{barrow2} J.D. Barrow, A.R. Liddle and C. Pahud, ``Intermediate inflation in light of the three-year WMAP observations'', {\it Physical Review D}, vol. 74, 127305, 2006.

\bibitem{barrow3} J.D. Barrow and N.J. Nunes, ``Dynamics of “logamediate” inflation'', {\it Physical Review D}, vol. 76, 043501, 2007.

\bibitem{Herrera} R. Herrera, N. Videla and M. Olivares, ``G-Warm inflation: Intermediate model'', arXiv:1811.05510v1 [gr-qc]. 
 
\bibitem{wmap} G. Hinshaw et al., ``Nine-year Wilkinson Microwave Anisotropy Probe (WMAP) observations: Cosmological parameter results'', {\it The Astrophysical Journal Supplement Series}, (25pp) vol. 208, 19, 2013.

\bibitem{planck1} P.A.R. Ade et. al. (Planck Collaboration), ``Planck 2013 results. I. Overview of products and scientific results'', {\it Astronomy and Astrophysics}, vol. 571, A22, 2014.

\bibitem{planck2} P.A.R. Ade et al. (Keck Array and BICEP2 Collaborations), ``Improved Constraints on Cosmology and Foregrounds from BICEP2 and Keck Array Cosmic Microwave Background Data with Inclusion of 95 GHz Band'', {\it Physical Review Letters}, vol. 116, 031302, 2016.

\bibitem{starobinsky1} A.A. Starobinsky, ``Spectrum of relict gravitational radiation and the early state of the universe'', {\it JETP Letters}, vol. 30, p. 682, 1979.

\bibitem{starobinsky2} A.A. Starobinsky, ``Cosmic background anisotropy induced by isotropic, flat-spectrum gravitational-wave perturbations'', {\it Soviet Astronomy Letters}, vol. 11, p. 133, 1985.


\end{thebibliography}
\end{document}